\documentclass{article}
\usepackage{spconf,amsmath,graphicx}

\title{PHASE-BASED INFORMATION FOR VOICE PATHOLOGY DETECTION}
\name{Thomas Drugman \thanks{Thomas Drugman is supported by the ``Fonds National de la Recherche
Scientifique'' (FNRS). The authors would like to thank Wallonia, Belgium, for its support (grant WALEO II ECLIPSE $\#516009$).}, Thomas Dubuisson, Thierry Dutoit}
\address{TCTS Lab - University of Mons - Belgium}
\begin{document}
%
\maketitle
\begin{abstract}
In most current approaches of speech processing, information is extracted from the magnitude spectrum. However recent perceptual studies have underlined the importance of the phase component. The goal of this paper is to investigate the potential of using phase-based features for automatically detecting voice disorders. It is shown that group delay functions are appropriate for characterizing irregularities in the phonation. Besides the respect of the mixed-phase model of speech is discussed. The proposed phase-based features are evaluated and compared to other parameters derived from the magnitude spectrum. Both streams are shown to be interestingly complementary. Furthermore phase-based features turn out to convey a great amount of relevant information, leading to high discrimination performance.
\end{abstract}
%

\section{Introduction}\label{sec:intro}

In the majority of current speech processing systems, information is captured from the amplitude component of the Fourier transform. However several studies on speech perception, such as \cite{Paliwal}, have highlighted the importance of phase information. In this way, phase-based features have recently shown their efficiency in various fields of speech processing, such as automatic speaker \cite{Murty} or speech \cite{Hegde} recognition. Among others, conclusions drawn in these works underline the complementarity of phase-based features with usual parameters extracted from the magnitude spectrum.

The main goal of this paper is precisely to investigate the potential of using features derived from the phase information for the application of automatic voice pathology detection, which, to the best of our knowledge, was never explored in the literature. For this, the paper is structured as follows. Section \ref{sec:PhaseFeat} details the phase-based features that are proposed. Relying on these features, Section \ref{sec:Exp} evaluates the efficiency of the phase information for dectecting voice disorders. Finally Section \ref{sec:conclu} concludes.

\section{Phase-based Features}\label{sec:PhaseFeat}

This Section presents the characteristics based on the speech signal phase information that are proposed in this paper. Analysis relying on the group delay function is first detailed in Section \ref{ssec:GD}. The respect of the mixed-phase model in both normo and dysphonic voices is then discussed in Section \ref{ssec:MixedPhase}.

\subsection{Group Delay-based Analysis}\label{ssec:GD}

\begin{figure*}[!ht]
  \centering
  \includegraphics[width=0.95\textwidth]{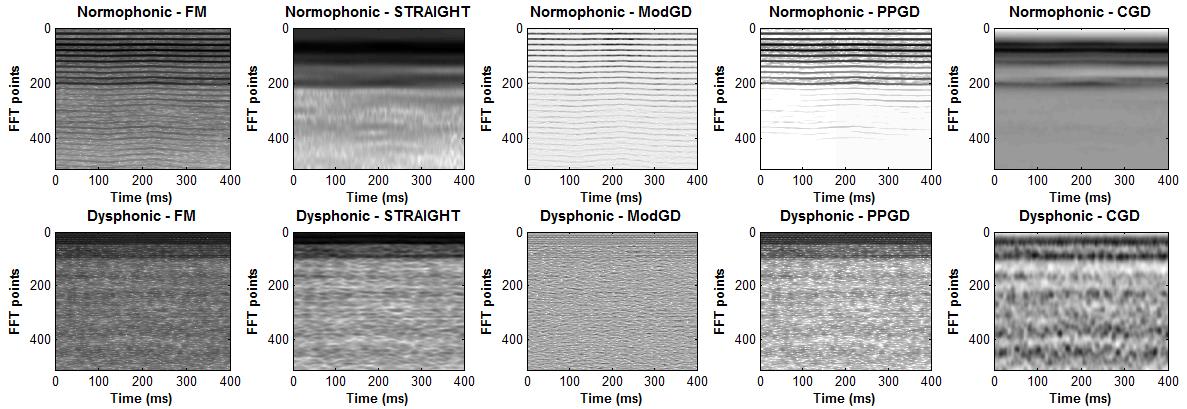}
  \caption{Illustration of the five types of spectrograms for a segment of sustained vowel $/a/$ for both normophonic (top plots) and dysphonic (bottom plots) voices. The spectrograms are related to the following representation (from left to right): the Fourier Magnitude (FM), the STRAIGHT spectrogram, the Modified Group Delay (ModGD), the Product of the Power and Group Delay (PPGD), and the Chirp Group Delay (CGD).}
  \label{fig:5Spec}
\end{figure*}

The group delay function is defined as the derivative of the unwrapped phase spectrum. However, the group delay computed in this way contains spikes due to the presence of some zeros of the signal z-transform close to the unit circle (where the Fourier transform is evaluated). Therefore group delay processing has been avoided for a long time. Nevertheless, some new representations aiming at reducing the effect of these spikes have been recently suggested, leading to some improvements in speech recognition, especially in noisy conditions \cite{Hegde}, \cite{PS}, \cite{Bozkurt-ChirpGD}. This section explores the use of these new modes of representation for determining the presence of a voice disorder. For this, 5 types of spectrograms are considered in this work:

\emph{i)} The \textbf{Fourier Magnitude (FM) spectrogram} is the commonly adopted representation in speech processing, here introduced as a baseline. \emph{ii)} The \textbf{STRAIGHT spectrogram} \cite{STRAIGHT} is based on a restructuration of the speech representation by using a pitch-adaptive time-frequency smoothing of the FM spectrogram. \emph{iii)} The \textbf{Modified Group Delay (ModGD) spectrogram} \cite{Hegde} is a function using a cepstral smoothing in order to reduce the effect of the spikes on the group delay function.
\emph{iv)} The \textbf{Product of the Power and Group Delay (PPGD) spectrogram} \cite{PS} is defined as the product of the power spectrum and the group delay function, aiming at reducing the source of spikes. \emph{v)} Finally, the \textbf{Chirp Group Delay (CGD) spectrogram} is a representation proposed by Bozkurt et al. in \cite{Bozkurt-ChirpGD} which relies on a chirp (i.e the Fourier transform is evaluated on a contour in the z-plane different from the unit circle) analysis of the zero-phase version of the speech signal. This approach was shown to provide a high-resolved representation of the formant peaks.

As an illustration, Fig. \ref{fig:5Spec} compares the five spectrograms for a segment of a normophonic and a dysphonic sustained vowel \emph{/a/}. It can be noticed from these plots that spectrograms of the normophonic voice present a structure well regular in time, while their equivalents for the dysphonic speech contain time-varying irregularities. These latter are probably due to the difficulties of the patient suffering from a voice disorder in sustaining a regular phonation. Indeed, during the production of a sustained vowel, the vocal tract can be assumed as a stationary system on a short-time period, even for pathological voices. The speech signal then results from the excitation of this stationary system by the glottal source. As a consequence, a regular glottal flow will be characterized by a smooth spectrogram. On the opposite, if the resolution of the spectrogram is sufficiently high, turbulences or cycle-to-cycle variations present in the glottal flow will be reflected by irregularities in the spectrogram structure, as exhibited in the bottom plots of Fig. \ref{fig:5Spec}. It is also observed in this figure that these irregularities are particularly well emphasized in the CGD spectrogram. This approach is indeed known for giving both a smooth and high-resolved representation for showing up resonance peaks in the speech spectrum \cite{Bozkurt-ChirpGD}.


\subsection{Respect of the Mixed-Phase Model of Speech}\label{ssec:MixedPhase}

According to the mixed-phase model \cite{MixedPhase}, voiced speech is composed of both minimum-phase (i.e causal) and maximum-phase (i.e anticausal) components. While the vocal tract and the glottal \emph{return phase} can be considered as minimum-phase systems, it has been shown \cite{MixedPhase} that the glottal \emph{open phase} is a maximum-phase signal. The key idea of the mixed-phase (or causal-anticausal) decomposition is then to separate both minimum and maximum-phase components of speech, where the latter is only due to the glottal contribution. By isolating the anticausal component of speech, the mixed-phase separation then allows to estimate the glottal open phase.

\begin{figure}[!ht]
  \centering
  \includegraphics[width=0.35\textwidth]{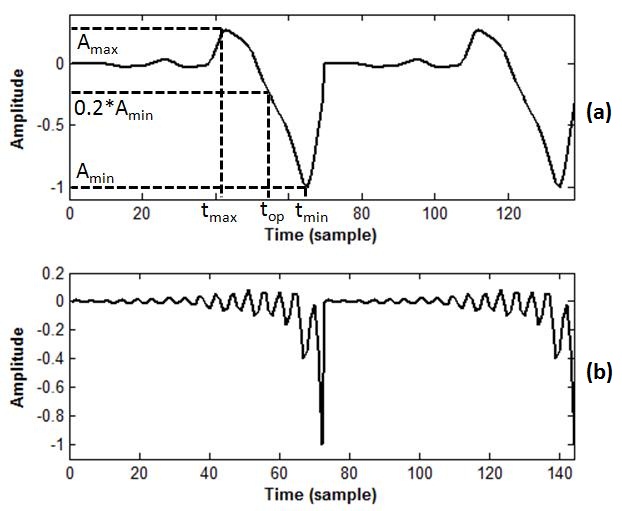}
  \caption{Two cycles of the anticausal component isolated by the mixed-phase decomposition \emph(a): when the speech segment exhibits characteristics of the mixed-phase model, \emph(b): when this is not the case. The particular instants defining the two proposed time constants are also indicated.}
  \label{fig:Raises}
\end{figure}

In this work, the mixed-phase deconvolution is achieved using the Complex Cepstrum-based Decomposition (CCD) proposed in \cite{Drugman-CCD}. If CCD is applied to a windowed segment of voiced speech exhibiting characteristics of the mixed-phase model, the source-tract separation can be correctly carried out. Two cycles of the resulting anticausal component are displayed in Fig. \ref{fig:Raises}(a), providing a reliable estimation of the glottal source (i.e corroborating the glottal flow models). If this is not the case, the decomposition fails and the resulting anticausal contribution has an irrelevant shape (as shown in Fig. \ref{fig:Raises}(b)), generally characterized by a high-frequency noise.

In order to assess the quality of the mixed-phase separation, two time features are extracted from the maximum-phase signal (see Figure \ref{fig:Raises}).  If $T_0$ denotes the pitch period, the first time constant $T_1$ is defined as $\frac{t_{min}-t_{max}}{T_0}$, while the second one $T_2$ is computed as $\frac{t_{min}-t_{op}}{T_0}$. These parameters should then contain relevant information about the consistency of the decomposition, i.e whether this deconvolution leads to a reliable estimation of the glottal flow or not.

\section{Evaluation of the Proposed Phase-based Features}\label{sec:Exp}

\subsection{Speech Material and Methods}

The speech corpus used in this study is the MEEI Disordered Voice Database \cite{Kay}, popular in the field of speech pathologies analysis. In this study, experiments are performed on sustained vowels uttered by 53 normophonic subjects and 657 subjects with a large panel of voice pathologies. Recordings resampled at 16 kHz are considered.

The five spectrograms detailed in Section \ref{ssec:GD} are computed using Blackman-windowed frames shifted every $10 ms$ and whose length is $30 ms$. For this, the Fourier spectrum is estimated by a DFT of 1024 points. All other parameters are fixed to the values recommended in the corresponding references. For each spectrogram, the relative difference between two consecutive frames is calculated, since this feature should convey relevant information about glottal source irregularities. In the rest of the paper, the prefix $d$ ahead of the name of a spectrogram type will be used for denoting this feature. It is worth emphasizing that since we consider frame-to-frame variations, the resulting features are robust to differences between recording conditions or equipments, and among others to phase distortion.

The mixed-phase decomposition is achieved using the CCD algorithm proposed in \cite{Drugman-CCD}. Since this method requires a GCI-synchronous process, Glottal Closure Instants (GCIs) are located on the speech signals using the DYPSA algorithm \cite{Dypsa}. Therefore the possible failure of GCI estimation for pathological voices is implicitly captured in the mixed-phase based features. For each resulting frame, one cycle of the anticausal component of speech is isolated and the two time constants $T_1$ and $T_2$ are extracted from it. So as to be synchronous with the other features, these streams are then interpolated at $100 Hz$. Besides the three spectral balances $Bal_1$, $Bal_2$ and $Bal_3$ proposed in \cite{Drugman-PathoIS09} are also extracted from the FM spectrogram. These balances are defined as the power spectral density in three perceptual subbands, and were shown to be highly discriminant for detecting a voice disorder.

\subsection{Mutual Information-based Evaluation}\label{ssec:MI}

In this part, the proposed features are compared according to their relevance for the problem of voice pathology detection. For this, we make use of information-theoretic measures \cite{Drugman-PathoIS09}, advantageous since it is independent of any classifier and allows an intuitive interpretation in terms of discrimination power. More precisely, the normalized Mutual Information (MI) \cite{Drugman-PathoIS09} of the proposed features is here studied. This measure is defined as the mutual information between the considered feature and the class labels (measure of the useful information), normalized over the entropy of the class labels (measure of the total class uncertainty). It can then be interpreted as the percentage of relevant information conveyed by the considered feature.

Table \ref{tab:TabMI} presents the values of the normalized MI for the 10 features. As expected from the results in \cite{Drugman-PathoIS09}, the two first spectral balances are strongly informative. From the various spectrogram representations, CGD provides the highest amount of relevant information (covering $56\%$ of the total uncertainty). Although in a lesser extent, ModGD and the two time constants characterizing the mixed-phase decomposition show an interesting potential for voice pathology detection. However, it is worth noting that the normalized MI is a measure of the intrinsic discrimination power of each feature separately, and consequently is not informative about the redundancy between them. Due to space limitation, the mutual information-based measures of redundancy are not exhaustively presented here. Nevertheless, the main conclusions that can be drawn from them are the following. Albeit spectral balances have the highest normalized MI in Table \ref{tab:TabMI}, they are also highly redundant. In this way, the joint use of $Bal_1$ and $Bal_2$ only brings $64.65\%$ of MI. On the contrary, the best combination of two features is surprisingly $T_2$ and $Bal_1$ (which is not straightforward at the only sight of Table \ref{tab:TabMI}), with $79.31\%$ of MI. This is possible as these two features present a very low amount of redundancy.

\begin{table*}[!htbp]
\centering
\begin{tabular}{| c || c | c | c | c | c | c | c | c | c | c |}
\hline
\textbf{Feature} & dFM & dSTRAIGHT & dModGD & dPPGD & dCGD & $T_1$ & $T_2$ & $Bal_1$ & $Bal_2$ & $Bal_3$ \\
\hline
\textbf{Normalized MI (\%)} & 22.32 & 16.32 & 30.56 & 15.43 & 55.97 & 32.02 & 23.09 & 56.64 & 55.85 & 15.39\\
\hline
\end{tabular}
\caption{Values of the normalized mutual information for the 10 features.}
\label{tab:TabMI}
\end{table*}



%

\subsection{Classifier-based Evaluation}

This section aims at evaluating the proposed features using a classifier for the automatic detection of voice disorders. For this, an Artificial Neural Network (ANN) is used for its discriminant learning capabilities. The ANN employed in this study has one hidden layer and uses sigmoid activation functions. The hidden layer is made of 16 neurons, as this gave the best trade-off between complexity and generalization capabilities in our attempts. Evaluation is achieved using a 10-fold cross validation framework. The system is finally assessed at both frame and patient levels, a patient being diagnosed as dysphonic if the majority of his frames are recognized as pathological.

\begin{table}[!ht]
\centering
\begin{tabular}{c | c | c}
\hline
\textbf{Features used} & \textbf{Error rate} & \textbf{Error rate}\\
& \textbf{(frame level)} & \textbf{(patient level)}\\
\hline
\hline
dFM & 17.2 & 8.73\\
\hline
dCGD & 9.40 & 4.93\\
\hline
$T_1$,$T_2$ & 13.28 & 5.35\\
\hline
$Bal_1$,$T_2$ & 8.65 & 5.07\\
\hline
$Bal_1$,$Bal_2$,$Bal_{3}$ & 9.97 & 7.89\\
\hline
ModGD,dPPGD,dCGD & 7.92 & 4.08\\
\hline
dFM,dSTRAIGHT, & 8.25 & 4.65\\
  dModGD,dPPGD,dCGD &   &  \\
\hline
$T_1$,$T_2$, & 7.97 & 4.08\\
 dModGD,dPPGD,dCGD &  &  \\
\hline
All 10 features & 6.16 & 4.08\\
\hline
\end{tabular}
\caption{Results of voice pathology detection using an ANN classifier for various feature sets.}
\label{tab:TabANN}
\end{table}

Results we obtained are presented in Table \ref{tab:TabANN} for different feature sets. Several conclusions can be drawn from these results. First of all, the efficiency of dCGD is confirmed as this feature alone leads to only $4.93\%$ of patients incorrectly classified. Its advantage over dFM spectrum can be noted. In the experiments using only 2 or 3 parameters, the improvement brought by the proposed features compared to the spectral balances is clearly seen. Indeed the two time constants characterizing the respect of the mixed-phase model lead, at the patient level, to a better classification than when using the 3 spectral balances. In addition, the use of $T_2$ in combination with $Bal_1$ is more performant than the 3 spectral balances, confirming the discussion about redundancy in Section \ref{ssec:MI}. It is worth noting the high performance achieved when using the 3 GD-based features. Adding the 2 magnitude-based spectrograms to these latter even leads to a slight degradation of accuracy. Similarly, it can be observed that adding the two mixed-phase model-based time constants makes the performance almost unchanged. Finally, considering all 10 features correctly identifies $93.84\%$ of frames and $95.92\%$ of patients. 
Although not reported in Table \ref{tab:TabANN}, it is worth noting that for the 3 GD-based features, the rates of false positive and negative patients are respectively of $16.98\%$ and $3.04\%$. The high rate of false positive detections can be explained by the unbalance of the MEEI database, leading to an overestimation of pathologies. Nonetheless, relying on a ROC curve, one could modify these latter rates by playing on the posterior threshold for deciding whether a frame is pathological or not (i.e a frame could be pathological with a probability greater or lower than 0.5).

\section{Conclusion}\label{sec:conclu}
This paper explored the potential of using phase-based features for detecting voice pathologies. It was shown that representations based on group delay functions are particularly suited for capturing irregularities in the speech signal. The respect of the mixed-phase model during the voice production was discussed and shown to convey relevant information. Besides it was underlined that phase-based and magnitude spectrum-based features may present interesting complementarity for this task, showing among others a very weak redundancy. The efficiency of these phase-based features may be explained by their higher sensitivity to turbulences during the phonation process.

\bibliographystyle{IEEEbib}
\bibliography{strings,refs}

\end{document}